\documentclass[preprint,showpacs,preprintnumbers,amsmath,amssymb]{revtex4-1}

\usepackage{graphicx,color}
\usepackage{dcolumn}
\usepackage{bm}

\begin{document}

\title{
Hamilton-Jacobi method for molecular distribution function
in a chemical oscillator
}

\author{Hiizu Nakanishi}
\affiliation{
Department of Physics, Kyushu University 33, Fukuoka 812-8581, Japan}

\author{Jun'ichi Wakou}
\affiliation{
Miyakonojo National College of Technology, Miyakonojo-shi, Miyazaki, 885-8567, Japan.
}

\author{Takahiro Sakaue}
\affiliation{
Department of Physics, Kyushu University 33, Fukuoka 812-8581, Japan}

\date{\today}

\begin{abstract}
Using the Hamilton-Jacobi method, we solve chemical Fokker-Planck
equations within the Gaussian approximation and obtain a simple and
compact formula for a conditional probability distribution.  The formula
holds in general transient situations, and can be applied not only for a
steady state but also for a oscillatory state.
By analyzing the long time behavior of the solution in the oscillatory
case, we obtain the phase diffusion constant along the periodic
orbit and the steady distribution perpendicular to it.
A simple method for numerical evaluation of these formulas are devised,
and they are compared with Monte Carlo simulations in the case of
Brusselator as an example.
Some results are shown to be identical to 
previously obtained expressions.
\end{abstract}
\pacs{83.80.Hj,83.60.Rs, 83.10.Ff,83.60.Wc}


\maketitle
\section{Introduction}

Chemical reactions are molecular processes and subject to the molecular
fluctuations.  Such fluctuations are negligible in homogeneous
macroscopic systems and the rate equation description is precise enough.
However, as the number of molecules involved decreases, the fluctuations
become visible and the system behavior may be quite different from the
one expected from the rate equation, especially in non-equilibrium
systems that show chemical oscillation, bifurcation, bistability, etc.

The study on stochasticity in chemical reactions has a long history as a
fundamental problem of irreversible statistical
physics\cite{Gardiner,Risken,vanKampen}, but its interests remained
largely academic until recently because experimental observations were
limited.
However, recent development of experiments by single molecule
measurement on biological systems starts revealing molecular
fluctuations in chemical reactions with extremely small numbers of
molecules\cite{Barkai-2000,Gonze-2002-a,Hirata-2002,Geva-2006,
Elowitz-2000,Atkinson-2003,Tigges-2009,Fung-2005,Nakanishi-2013}.
This should provide abundant observations on molecular fluctuations in
chemical reactions, and will shed new light on this old problem.

The molecular fluctuations in chemical reactions have been studied by
chemical master equations for the distribution function of molecular
numbers of each chemical species\cite{Gardiner,Risken,vanKampen}.
For spatially homogeneous systems, transition probabilities of molecular
reactions are given by a function of molecular concentrations with a
coefficient proportional to the system size $\Omega$.
In the large $\Omega$ limit, Kramers-Moyal
expansion\cite{Kramers-1940,Moyal-1949} of the chemical master equation
gives the Fokker-Planck equation, where the molecular fluctuation
effects are represented by the generalized diffusion term of the order
of $1/\Omega$.
Using the analogy between the Fokker-Planck equation and the
Schr\"odinger equation, the Hamilton-Jacobi formalism has been employed
to obtain the distribution function in terms of the potential
function\cite{Kubo-73,Kitahara-1975}, and
there have been  attempts to solve this equations using the
$1/\Omega$ expansion\cite{Tomita-1974a, Tomita-1974b, Vance-1996} and
the Mori-Zwanzig projection method\cite{Grossmann-1978,Schranner-1979}.

Gaspard has developed the
method to calculate the correlation time of chemical
oscillators\cite{Gaspard-2002a,Gaspard-2002b,Gonze-2002}.
With his formalism, the probability distribution for the molecular
numbers can be obtained for a non-stationary state by solving
corresponding Hamilton's canonical equations.
In this work, we further develop the formalism to obtain 
an explicit expression for the
distribution functions in general situations including the case of
chemical oscillations.

%

%

The paper is organized as follows.  In Sec. II, we review the
Hamilton-Jacobi formalism for the chemical master equation to establish
the notations, then in Sec. III our main result is derived, i.e. the
fundamental solution for the Fokker-Planck equation in the Gaussian
approximation.  In Sec. IV, a simple method for numerical evaluation of
the formula is devised, and the formula is examined in the situation of
a steady state and that of a oscillatory state. It is shown that our
formula reduces to the one by the linear noise approximation in the case
of steady states.  In Sec. V, taking the Brusselator as an example, the
formulas are evaluated numerically and compared with the results of Monte
Carlo simulations.  The concluding remarks are given in Sec. VI.
Detailed derivations of some of the results are presented in Appendices.

\section{Formulation}
We review the Hamilton-Jacobi method for the chemical master equation to
establish the notation,  basically following the formalism by
Gaspard\cite{Gaspard-2002b}.

\subsection{Molecular fluctuation in chemical reactions}

Consider the system that contains $d$ chemical species denoted by
${\rm X}_i$ ($i=1,2,\cdots, d$), among which
$r$ chemical reactions are taking place:
\begin{equation}
\sum_{i=1}^d \nu^i_{\rho-}{\rm X}_i \quad\rightarrow\quad
\sum_{i=1}^d \nu^i_{\rho+}{\rm X}_i 
\quad (\rho=1, 2, \cdots, r),
\label{chem-reactions}
\end{equation}
where $\nu^i_{\rho\pm}$ are integers.  Thus, the number of $i$'th
molecule $X_i$ changes by the $\rho$'th reaction event as
\begin{equation}
X_i \quad\rightarrow\quad X_i+\Delta X^i_\rho
\end{equation}
with the stoichiometric coefficient
$\Delta X^i_\rho\equiv \nu^i_{\rho+}-\nu^i_{\rho-}$.
%
In order that a steady chemical oscillation be possible, the system
needs steady input and output flows, which may also be included in
Eq. (\ref{chem-reactions}) as the form of chemical reactions.

Let $\Omega$ be the system volume.
Then the transition rate $W_\rho$ of the reaction $\rho$ may be given by
\begin{equation}
W_\rho (\bm X) =
  k_\rho \Omega \prod_{i=1}^d \prod_{m=1}^{\nu^i_{\rho-}}
    {X_i - m+1\over\Omega},
\end{equation}
with the reaction constant $k_\rho$, assuming the system is
spatially homogeneous.
Then, the master equation for the probability distribution of the
molecular numbers $P(\bm X,t)$
is given by
\begin{equation}
{d\over dt}P(\bm X,t) = \sum_{\rho=1}^r
\Bigl[
W_\rho(\bm X-\Delta\bm X_\rho) P(\bm X-\Delta\bm X_\rho, t) -
W_\rho(\bm X) P(\bm X,t)
\Bigr].
\label{master-eq}
\end{equation}

\subsection{Chemical Fokker-Planck equation}

In the large $\Omega$ limit, this can be approximated as the chemical
Fokker-Planck equation for the concentration distribution function
\begin{equation}
P(\bm x,t) \equiv \Omega^{-d} P(\bm X,t);\quad
\bm x \equiv {\bm X\over\Omega}.
\end{equation}
In the lowest order of $\Omega$, the Fokker-Planck equation 
may be written in the form of
\begin{equation}
{1\over\Omega}{\partial\over\partial t}P(\bm x,t)
= -H\left(\bm x, {1\over\Omega}{\partial\over\partial \bm x}\right)
P(\bm x,t)
\label{FP-eq}
\end{equation}
using the ``Hamiltonian''
\begin{equation}
H(\bm x, \bm p) \equiv 
\sum_i p_i F_i(\bm x) - \sum_{ij}p_i p_j Q_{ij}(\bm x),
\label{Hamiltonian}
\end{equation}
where
\begin{eqnarray}
F_i(\bm x) & \equiv & 
\lim_{\Omega\to\infty}
{1\over\Omega}\sum_{\rho=1}^r \Delta X_\rho^i\, W_\rho(\Omega \bm x) ,
\label{F_i}
\\
Q_{ij}(\bm x) & \equiv &
\lim_{\Omega\to\infty}
{1\over 2}{1\over\Omega}\sum_{\rho=1}^r 
\Delta X_\rho^i\Delta X_\rho^j\, W_\rho(\Omega \bm x).
\label{Q_ij}
\end{eqnarray}
The $F$-term represents the average flow given by the rate equations
\begin{equation}
{d\over dt}\bm x = \bm F(\bm x),
\label{rate-eq}
\end{equation}
while the $Q$-term represents the diffusion due to the molecular
fluctuation around the average motion.

\subsection{Hamilton-Jacobi method}

Assuming  the form of solution for Eq.(\ref{FP-eq}) as
\begin{equation}
P(\bm x, t) = \exp\left[\Omega\phi(\bm x,t)\right]
\label{P(x,t)}
\end{equation}
with the function $\phi(\bf x, t)$,
we can put the Fokker-Planck equation in the form of
Hamilton-Jacobi equation
\begin{equation}
{\partial\phi(\bm x, t)\over\partial t}
+ H\left(\bm x, {\partial \phi(\bm x, t)\over\partial t}\right)=0
\label{HJ-eq}
\end{equation}
in the lowest order of $1/\Omega$.

According to
Hamilton-Jacobi theory of classical mechanics,
the solution $\phi(\bm x,t)$ of Eq.(\ref{HJ-eq}) gives
the position
$\bm x(t)$ and the momentum
\begin{equation}
\bm p(t) = {\partial \phi(\bm x, t)\over\partial \bm x}
\label{p(t)}
\end{equation} 
that satisfy Hamilton's canonical equations of motion
\begin{eqnarray}
\dot x_i(t) & = & {\partial H\over\partial p_i}
= F_i(\bm x) - \sum_j 2Q_{ij}(\bm x)p_j,
\label{H-eq-x}
\\
\dot p_i(t) & = & -{\partial H\over\partial x_i}
= -\sum_j {\partial F_j(\bm x)\over\partial x_i} p_j
 + \sum_{j,k}{\partial Q_{jk}(\bm x)\over\partial x_i}p_j p_k .
\label{H-eq-p}
\end{eqnarray}
Conversely, the solution of Hamilton-Jacobi equation (\ref{HJ-eq}) can
be constructed through $\bm x(t)$ and $\bm p(t)$ that satisfy Hamilton's
canonical equations (\ref{H-eq-x}) and (\ref{H-eq-p}) as
\begin{eqnarray}
\phi(\bm x, t) & = & \int_0^t\left(
\sum_i p_i(t'){dx_i(t')\over dt'}
    -H\bigl(\bm x(t'), \bm p(t')\bigr)\right)dt'   + \phi_0(\bm x_0)
\nonumber\\
& \equiv &
J(\bm x_0,\bm p_0,t)   + \phi_0(\bm x_0),
\label{phi}
\end{eqnarray}
where $\phi_0(\bm x)$ is the initial condition of $\phi(\bm x, t)$ at
$t=0$, and $\bm x_0$ and $\bm p_0=\partial\phi_0/\partial \bm x_0$ are
the initial values of $\bm x(t)$ and $\bm p(t)$ that lead to $\bm x$ at
the time $t$;
The function $J(\bm x_0,\bm p_0,t)$ is the action integral along $\bm
x(t)$ and $\bm p(t)$ with this initial condition.  
Note that the rate equation (\ref{rate-eq}) is given by the canonical
equation (\ref{H-eq-x}) within the $\bm p=0$ subspace, where $H=0$.

\section{Approximate solution of Hamilton-Jacobi equation}

We will derive an approximate solution for $\phi(\bm x, t)$ to obtain a
non-equilibrium distribution function (\ref{P(x,t)}) in the Gaussian
approximation.  Suppose that the system starts from a point $\bm x^*_0$
at time $t=0$.
We take
the initial distribution to be Gaussian given by
\begin{equation}
\phi(\bm x, 0) = \phi_0(\bm x)
 = -{1\over 2\sigma_0^2}(\bm x- \bm x^*_0)^2,
\label{phi_0}
\end{equation}
and will take the infinitesimal limit of
$\sigma_0^2$ after we obtain the expression for $\phi(\bm x, t)$.

Let $\bm x^*(t)$ be the solution of the rate equation (\ref{rate-eq})
with the initial condition $\bm x^*(0)=\bm x_0^*$, then
$\bm x^*(t)$ and $\bm p^*(t)=0$ are the solution of Hamilton's
canonical equations (\ref{H-eq-x}) and (\ref{H-eq-p}) with the initial
condition
\begin{equation}
\bm x^*(0) = \bm x^*_0, \qquad
\bm p^*(0) = 
{\partial\phi(\bm x, 0)\over\partial\bm x}\Bigr|_{\bm x=\bm x_0^*}=0.
\label{x*0_p*0}
\end{equation}
One can see that
the maximum of the distribution is always located at $\bm x^*(t)$
because
\begin{equation}
{\partial \phi(\bm x, t)\over \partial \bm x}
\Bigr|_{\bm x=\bm x^*(t)} = \bm p^*(t) = 0.
\end{equation}
%
We expand $\phi(\bm x,t)$ around this maximum point $\bm x=\bm x^*(t)$
as
\begin{eqnarray}
\phi(\bm x, t) & \approx &
\phi(\bm x^*(t),t)
+{1\over 2}\sum_{i,j}{\partial^2\phi\over\partial x_i\partial x_j}\biggr|_*
(x_i-x^*_i(t))(x_j-x^*_j(t))
\nonumber\\
& \equiv & -\,{1\over 2}\sum_{i,j}\hat M^{-1}_{ij}(t)(x_i-x^*_i(t))(x_j-x^*_j(t)),
\label{phi-expansion}
\end{eqnarray}
where $|_*$ denotes that the derivative is evaluated 
at $\bm x=\bm x^*(t)$.
Note that $\phi(\bm x^*,t)=0$ from Eq.(\ref{phi}) because $\bm
p^*=0$ thus $H(\bm x^*, \bm p^*)=0$.
Then, the distribution function is written as
\begin{equation}
P(\bm x, t| \bm x_0^*) = \left({\Omega\over 2\pi}\right)^{d/2}
{1\over\sqrt{|\hat M(t)|}} \exp\left[ -{\Omega\over 2}
\bigl(\bm x-\bm x^*(t)\bigr)^T\hat M(t)^{-1}
                  \bigl(\bm x-\bm x^*(t)\bigr) \right]
\label{P(x,t)-M}
\end{equation}
within Gaussian approximation,
and the variances are obtained by
\begin{equation}
\left<(x_i-x^*_i)(x_i-x^*_j)\right> \equiv 
\int d\bm x\, (x_i-x^*_i)(x_i-x^*_j) P(\bm x, t| \bm x_0^*)
= {1\over\Omega}\,\hat M_{ij}(t) ,
\end{equation}
using the matrix $\hat M$ defined by Eq.(\ref{phi-expansion}).

Note that Eq.(\ref{P(x,t)-M}) represents the conditional probability
distribution, i.e.  the fundamental solution of the Fokker-Planck
equation with the initial distribution localized at $\bm x=\bm x_0^*$.
We will derive a compact expression of the covariance matrix $\hat M(t)$
in the following.

\subsection{Expansion around the rate equation orbit $\bm x^*(t)$}

The expansion coefficient of Eq.(\ref{phi-expansion}) may be written as
\begin{equation}
-\hat M_{ij}^{-1}(t) 
=
{\partial^2\phi(\bm x, t)\over\partial x_i\partial x_j}\biggr|_*
=
{\partial p_j\bigl(\bm x_0(\bm x,t),t\bigr)\over\partial x_i}\biggr|_*,
\label{M-inv-def}
\end{equation}
where $p_j(\bm x_0, t)$ denotes the $j$-th component of the momentum at
$t$ from the initial condition $(\bm x_0, \bm p_0)$ with $\bm
p_0=\partial\phi_0/\partial\bm x|_{\bm x=\bm x_{0}}$.  Conversely, the
initial position $\bm x_0$ can be denoted by $\bm x_0(\bm x,t)$ as a
function of the ending point $\bm x$ at which the solution of Hamilton's
equations arrives at the time $t$.  Note that the momentum $\bm p$ is
always related to $\bm x$ by Eq.(\ref{p(t)}).

In order to evaluate Eq.(\ref{M-inv-def}), we will consider the linear
expansion of Hamilton's equations (\ref{H-eq-x}) and (\ref{H-eq-p})
around the rate equation orbit $(\bm x^*(t), 0)$: 
\begin{align}
\delta\dot{\bm x}(t) & = 
\hat L(t)\delta\bm x(t) - 2\hat Q(t)\delta\bm p(t),
\label{delta_x-eq}
\\
\delta\dot{\bm p}(t) & = -\hat L^\dagger(t)\delta\bm p(t),
\label{delta_p-eq}
\end{align}
where  $(\delta\bm x(t), \delta\bm p(t))\equiv(\bm x(t)-\bm
x^*(t), \bm p(t))$.
The matrix $\hat L(t)$ and $\hat Q(t)$ are defined as
\begin{equation}
L_{ij}(t)\equiv
   {\partial F_i\over\partial x_j}\biggr|_{\bm x=\bm x^*(t)},
\qquad
Q_{ij}(t)\equiv Q_{ij}(\bm x^*(t)),
\label{L_ij}
\end{equation}
and $\dagger$ denotes the transposed matrix.
Note that the operator $\hat L(t)$ represents the time development of
$\delta \bm x$ within the $\bm p=0$ subspace, namely the deviation given
within the rate equation (\ref{rate-eq}).

We define the time evolution operator for the deviation within the rate
equation, i.e. for $\delta\bm x(t)$ within the
$\delta\bm p=0$ subspace,
\begin{equation}
\hat U_L(t,t_0) \equiv 
{\cal T}\exp\left[\int_{t_0}^t dt' \hat L(t')\right],
\label{hat_U}
\end{equation}
where  ${\cal T}$ is the time ordering operator. 
Then the formal solution of
Eqs.(\ref{delta_x-eq}) and (\ref{delta_p-eq}) for both $\delta \bm x$
and $\delta\bm p$
 can be written down as
\begin{align}
\delta\bm x(t) & =
\hat U_L(t,0)\left[\delta\bm x(0) -2\hat{\cal Q}_L(t) \delta\bm p(0)
\right],
\label{delta_x-sol0}
\\
\delta\bm p(t) & = \hat U_L^\dagger(0,t)\delta\bm p(0),
\label{delta_p-sol0}
\end{align}
where we have introduced the symmetric matrix
\begin{equation}
\hat{\cal Q}_L(t) \equiv 
\int_0^t dt' \hat U_L(0, t')\hat Q(t')\hat U_L^\dagger(0,t').
\label{Q_L(t)}
\end{equation}
Basic properties of $\hat U_L(t,t_0)$ are given in Appendix A.

\subsection{Expression for $\hat M(t)$}

The initial deviation $\delta\bm p(0)$ is related to $\delta\bm x(0)$
through Eqs.(\ref{phi_0}) and (\ref{x*0_p*0}) as
\begin{equation}
\delta\bm p(0) 
= {\partial\phi_0(\bm x)\over\partial\bm x}\Bigr|_{\bm x=\bm x(0)}
= -{1\over\sigma_0^2}\delta\bm x(0),
\end{equation}
therefore, Eqs.(\ref{delta_x-sol0}) and (\ref{delta_p-sol0}) become
\begin{align}
\delta\bm x(t) & =
\hat U_L(t,0)\left[1 -2\hat{\cal Q}_L(t)\left(-{1\over\sigma_0^2}\right)
\right]\delta\bm x(0),
\label{delta_x-sol}
\\
\delta\bm p(t) & = 
\hat U_L^\dagger(0,t)\left(-{1\over\sigma_0^2}\right)\delta\bm x(0),
\label{delta_p-sol}
\end{align}
which lead to the relation between $\delta\bm p(t)$ 
and $\delta\bm x(t)$,
\begin{align}
\delta\bm p(t) &  =
\hat U_L^\dagger(0,t)\left(-{1\over\sigma_0^2}\right)
\left[1 -2\hat{\cal Q}_L(t)\left(-{1\over\sigma_0^2}\right)
\right]^{-1}
\hat U_L(0,t)\,\delta\bm x(t)
\nonumber\\
& \to
-\,\hat U_L^\dagger(0,t)\,
    {1\over 2}\hat{\cal Q}_L^{-1}(t)\,\hat U_L(0,t)\,\delta\bm x(t)
\end{align}
in the limit of $\sigma_0^2\to 0$.  Using Eq.(\ref{M-inv-def}),
we obtain the expression for the covariance matrix
\begin{equation}
\hat M^{-1}(t) =
\hat U_L^\dagger(0,t)\, {1\over 2}\hat{\cal Q}_L^{-1}(t)\,\hat U_L(0,t),
\label{hat_M-inv}
\end{equation}
or
\begin{equation}
\hat M(t) = 
\hat U_L(t,0)\, 2\hat{\cal Q}_L(t)\,\hat U_L^\dagger(t,0).
\label{hat_M}
\end{equation}
Note that the covariance matrix satisfies
\begin{equation}
{d\over dt} \hat M(t) =
\hat L(t)\hat M(t) + \hat M(t)\hat L^\dagger(t) + 2\hat Q(t)
\label{dM/dt}
\end{equation}
with the initial condition $\hat M(0)=0$.  This equation has been
obtained by the $1/\Omega$ expansion of fluctuation\cite{Tomita-1974a}.


\section{Distribution functions}

The major result of this paper is the expression of the covariance
matrix (\ref{hat_M}) for the probability distribution (\ref{P(x,t)-M}),
which holds in general transient situations as long as the Gaussian
approximation is valid.
In this section, first we discuss the numerical method to estimate our
formulas, then we will examine this result for two cases: the case where
the rate equation leads to a stationary state, and the case where the
rate equation gives rise to a stable oscillation.

\subsection{Numerical estimate of the formulas}

At this point,
it is convenient to introduce the bra and ket notation for 
the row  and  column vectors, with which
Eqs.(\ref{delta_x-sol0}) and (\ref{delta_p-sol0}) are 
expressed as
\begin{align}
\left|\delta x(t)\right> & =
\hat U_L(t,0)\left[\left|\delta x(0)\right> -
     2\hat{\cal Q}_L(t) \left|\delta p(0)\right>
            \right],
\label{ket-delta_x-sol}
\\
\left<\delta p(t)\right| & = 
     \left<\delta p(0)\right|\hat U_L(0,t).
\label{bra-delta_p-sol}
\end{align}

The components of the covariance matrix $\hat M(t)$ of
Eq.(\ref{hat_M}) and $\hat{\cal Q}_L(t)$ of Eq.(\ref{Q_L(t)}) can be
numerically estimated easily by solving these equations
with proper initial conditions.
Let us first introduce the following notations for the solutions of
Eqs.(\ref{ket-delta_x-sol}) and (\ref{bra-delta_p-sol}):
\begin{align}
\left|\delta_x x(t;\delta x_0)\right> & \equiv
\hat U_L(t,0)\left|\delta x_0\right> ,
\label{delta_r_x}
\\
\left|\delta_p x(t;\delta p_0)\right>  & \equiv
-\hat U_L(t,0) 2\hat{\cal Q}_L(t) \left|\delta p_0\right>,
\label{delta_p_x}
\\
\left<\delta p(t;\delta p_0)\right| & \equiv
     \left<\delta p_0\right|\hat U_L(0,t).
\label{delta_p_p}
\end{align}
Eq.(\ref{delta_r_x}) may be interpreted as the deviation of $\bm x$ from
$\bm x^*$ by the rate equation with the initial deviation $\delta \bm
x_0$ within the $\bm p=0$ subspace.  Eqs.(\ref{delta_p_x}) and
(\ref{delta_p_p}) are the deviation of $\bm x$ and $\bm p$,
respectively, with the initial deviation $\delta\bm p_0$.  They can be
obtained numerically by solving Eqs.(\ref{delta_x-eq}) and
(\ref{delta_p-eq}) with the initial conditions $(\delta\bm x,\delta\bm
p)=(\delta\bm x_0,0)$ or $(\delta\bm x,\delta\bm p)=(0, \delta\bm p_0)$
at $t=0$.

Let $\hat I$ be the identity matrix and this may be expressed as
\begin{equation}
\hat I = \sum_\alpha \left|\alpha\right> \left<\alpha\right|
\end{equation}
by a complete set of orthonormal base vectors $\left|\alpha\right>$.
Then, a matrix element $\left<i\right|\hat M(t)\left|j\right>$ of
Eq.(\ref{hat_M}) can be put in the form
\begin{align}
\left<i\right|\hat M(t)\left|j\right> & =
\sum_{\alpha} \left<i\right| \hat U_L(t,0)
 2\hat{\cal Q}_L(t) \left|\alpha\right> \left<\alpha\right| 
\hat U_L^\dagger(t,0)\left|j\right>
\nonumber \\
& =
\sum_{\alpha} -\left<i| \delta_p x(t;\alpha)\right> 
 \left<\delta_x x(t; \alpha)|j\right>,
\label{M_ij}
\end{align} 
where $\left<\delta_x x(t;\alpha)\right|$ is a transposed column vector
of $\left|\delta_x x(t;\alpha)\right>$.
Similarly, the matrix elements of $\hat{\cal Q}_L(t)$ can be
expressed as
\begin{align}
\left<i\right|2\hat{\cal Q}_L(t)\left|j\right>
& = 
\left<i\right|\hat U_L(0,t)\hat U_L(t,0)
  2\hat{\cal Q}_L(t)\left| j\right>
\nonumber \\
& = 
-\left<\delta p(t;i)|\delta_p x(t;j)\right>.
\label{QL_ij}
\end{align}
With these formulas, we can evaluate the matrix elements numerically
simply by solving the ODE's, i.e. Eqs.(\ref{delta_x-eq}) and
(\ref{delta_p-eq}) with proper initial conditions.

\subsection{Fluctuations around the steady state}

In the simple case where the rate equation leads to a stationary state
$\bm x_s^*$, we show that Eq.(\ref{hat_M}) results in the expressions
that have been obtained by the linear noise approximation for the
Fokker-Planck equation.

In this case, the matrices $\hat L(t)$ and $\hat Q(t)$ of
Eq.(\ref{L_ij}) converge to time-independent ones as the rate equation
solution $\bm x^*(t)$ does,
\begin{equation}
 \lim_{t\to\infty} \bm  x^*(t) = \bm x^*_s,\quad
 \lim_{t\to\infty} \hat L(t) \equiv \hat L_s,\quad
 \lim_{t\to\infty} \hat Q(t) \equiv \hat Q_s,
\end{equation}
therefore,
the fluctuation distribution $P_s(\bm x)$
around the steady state is given by
\begin{equation}
P_s(\bm x) = \left({\Omega\over 2\pi}\right)^{d/2}
{1\over\sqrt{|\hat M_s|}} \exp\left[ -{\Omega\over 2}
\bigl(\bm x-\bm x_s^*\bigr)^T\hat M_s^{-1}
                  \bigl(\bm x-\bm x_s^*\bigr) \right]
\label{P_s(x)}
\end{equation}
with
the covariance matrix $\hat M_s$ determined by
\begin{equation}
\hat L_s\hat M_s + \hat M_s\hat L_s^\dagger + 2\hat Q_s=0
\end{equation}
because the covariance matrix should satisfies Eq.(\ref{dM/dt}). 
The time correlation of fluctuation around the steady state is obtained
as
\begin{align}
\lefteqn{
 \left< \left(\bm x(t)-\bm x_s^*\right)\left(\bm x(0)-\bm x_s^*\right)^T
\right>
}
\nonumber\\
& =
\int\left[
\left(
    \int \left(\bm x_t-\bm x_s^*\right)P(\bm x_t,t|\bm x_0)\, d\bm x_t
\right)
\left(\bm x_0-\bm x_s^*\right)^T P_s(\bm x_0) \right] d\bm x_0
\nonumber\\
& =
\int
\left( \bm x^*(t)-\bm x_s^* \right)
\left(\bm x_0-\bm x_s^*\right)^T P_s(\bm x_0) \, d\bm x_0
\nonumber\\
& =
\int \left[e^{\hat L_s t}\left( \bm x_0-\bm x_s^* \right)\right]
\left(\bm x_0-\bm x_s^*\right)^T P_s(\bm x_0) \, d\bm x_0
=
{1\over\Omega}\, e^{\hat L_s t}\hat M_s,
\end{align} 
where $t>0$ and $\bm x^*(t)$ is the rate equation solution with the
initial condition $\bm x^*(0)=\bm x_0$.
We have used that
$\bm x^*(t)-\bm x_s^* =e^{\hat L_s t}\left( \bm x_0-\bm x_s^* \right)$,
assuming that $\bm x^*(t)$ is close to $\bm x_s^*$.
These expressions are identical to those obtained by the linear noise
approximation\cite{vanKampen,Elf-2003,Scott-2007}.

\subsection{Fluctuations around the periodic orbit}

With our formula (\ref{P(x,t)-M}) with Eq.(\ref{hat_M-inv}), we can also
study the fluctuations around a periodic oscillation.
In this subsection, we present analysis on the long time behavior of the
distribution around a stable oscillation.

In an autonomous oscillatory system, the distribution relaxes in two
ways: the phase diffusion along the periodic orbit and the relaxation
within the space perpendicular to the periodic orbit.
The distribution diffuses along the periodic orbit
because there is no restoring force due to the time translational
symmetry; the distribution spreads over the whole orbit after many
periods of time.
On the other hand, the distribution spreads perpendicular to the orbit
to reach a steady form relatively fast.  The width of the distribution
varies along the orbit, depending on the local stability of the orbit.
We study these changes by considering the covariance matrix $\hat M(t)$
after many times of the period.

\subsubsection{Time evolution operator $\hat U$
over the period}

Let $\bm x^*(t)$ be the periodic solution for the rate equation
(\ref{rate-eq}):
\begin{equation}
   \bm x^*(t+T) = \bm x^*(t),
\label{periodic_x^*}
\end{equation} 
where $T$ is the period.  The initial point on the periodic orbit is
denoted by $\bm x_0^*\equiv \bm x^*(0)$.
Now, we define the time evolution operator $\hat U$ around this orbit
over the period $T$,
\begin{equation}
\hat U \equiv \hat U_L(T,0).
\label{U}
\end{equation}
Note that $\hat U$ depends on the starting point $\bm x^*(0)\bm =\bm
x_0^*$.  In this work, we consider only the simplest case where $\hat U$
have neither degenerate eigenvalues nor Jordan blocks larger than 1.
We describe some of the properties of this operator in this subsection.

The $i$'th eigenvalue of $\hat U$ is denoted by $\lambda_i$ and its
right and left eigenvectors by $\left|e_i\right>$ and
$\left<f_i\right|$, respectively,
\begin{equation}
\hat U\left|e_i\right> = \lambda_i\left|e_i\right>,
\quad
\left<f_i\right| \hat U =\left< f_i\right|  \lambda_i,
\end{equation}
with the normalization
\begin{equation}
\left<f_i | e_j\right> = \delta_{i,j}.
\label{norm}
\end{equation}
For an autonomous system, the largest eigenvalue is 1 
and its  eigenvectors may be given by
\begin{equation}
 \lambda_1=1, \quad 
\left|e_1\right> = \left|F(\bm x_0^*)\right>, \quad
\left<f_1\right| = 
{\partial \left< p^*(0;E)\right|\over\partial E}\biggr|_{E=0}.
\end{equation}
Here, $\bm p^*(t;E)$ is the periodic orbit outside the $\bm p=0$
plane, where the value of the Hamiltonian (\ref{Hamiltonian}) is
non-zero, $E\ne 0$ (see Appendix B).  The absolute values of other
eigenvalues are smaller than 1 because the periodic orbit is stable.
For the simplest case we are considering now, $\hat U$ can be expressed
as
\begin{equation}
\hat U = \sum_i \left|e_i\right>\lambda_i\left<f_i\right|.
\label{U_spectral}
\end{equation} 

Now, we define
\begin{equation}
 \left|e_i(t)\right>\equiv\hat U_L(t,0)\left|e_i\right>,
\quad
 \left<f_i(t)\right|\equiv \left<f_i\right|\hat U_L(0,t),
\end{equation}
then it is easy to show that they are the right and the left
eigenvectors of the time evolution operator $\hat U_L(T+t,t)$ over the
period at $\bm x^*(t)$, i.e.,
\begin{equation}
 \hat U_L(T+t,t) \left|e_i(t)\right>=\lambda_i\left|e_i(t)\right>,
\quad
 \left<f_i(t)\right| \hat U_L(T+t,t)=\lambda_i \left<f_i(t)\right|,
\end{equation}
and
\begin{equation}
 \left<f_i(t)|e_j(t)\right> = \delta_{ij},
\end{equation}
with
\begin{equation}
 \left|e_i(T)\right> = \lambda_i\left|e_i\right>,
\quad
 \left<f_i(T)\right| = {1\over\lambda_i}\left<f_i\right|.
\end{equation}
With these notations, we have the expression
\begin{equation}
 \hat U_L(t,0) = \sum_{\ell}\left|e_\ell(t)\right>\left<f_\ell\right|.
\end{equation}

\subsubsection{Phase diffusion and period fluctuation}

Now, we can obtain the expression for the phase diffusion along the
periodic orbit.  The unit tangential vector $\hat{\bm n}_\parallel$ of
the periodic orbit at $\bm x^*(t)$ is given as a function of time by
\begin{equation}
\hat{\bm n}_\parallel(t) 
\equiv {\bm F(\bm x^*(t))\over |\bm F(\bm x^*(t))|}
\end{equation} 
Thus the spatial variance of the orbit
in the tangential direction $\left< \Delta x_\parallel^2(t)\right>$ is
obtained from
\begin{eqnarray}
\left< \Delta x_\parallel^2(t)\right>
&= &
{1\over\Omega}\,
\left<\hat n_\parallel(t)\right|\hat M(t)\left|\hat n_\parallel(t)\right>
\nonumber\\
&= &
 {1\over\Omega}\,{1\over  |\bm F(\bm x^*(t))|^2}
\left< F(\bm x^*(t))\right|\hat M(t)\left| F(\bm x^*(t)) \right>
\end{eqnarray}
in the braket notation.


Let $\left<\bigl(\Delta t_{\bm x^*(t)}\bigr)^2\right>$ be the variance
of the time when each sample passes through the plane perpendicular to
the orbit at $\bm x^*(t)$, then up to the lowest order in $\Omega$, this
can be estimated by $\left< \Delta x_\parallel^2(t)\right>$ divided by
the square of the average speed,
\begin{align}
\left<\bigl(\Delta t_{\bm x^*(t)}\bigr)^2\right>
& = 
{1\over |\bm F(\bm x^*(t))|^2}\,
\left< \Delta x_\parallel^2(t)\right>
\nonumber \\
& =
{1\over \Omega}\, {1\over |\bm F(\bm x^*(t))|^4}
\left<F(\bm x^*(t))\right| \hat M(t) \left| F(\bm x^*(t))\right>.
\end{align}

Using this expression, we define the period fluctuation 
 $\left<\Delta T^2\right>$  by
\begin{align}
\left<\Delta T^2\right> & \equiv 
\lim_{r\to\infty}{1\over r}\left<\Delta t_{\bm x^*(rT)}^2\right>
\label{Delta_T^2_0}
\end{align}
where $r$ is an integer.
We obtain the compact expression
\begin{equation}
\left<\Delta T^2\right> =
{1\over \Omega}\, \left<f_1\right| 2\hat{\cal Q}_L(T)\left|f_1\right>.
\label{Delta_T^2}
\end{equation} 
The detailed derivation is given in Appendix C.
%

\subsubsection{Distribution perpendicular to the periodic orbit}

Let $\hat{\bm n}_{\perp,i}(t)$ ($2\le i \le d$) be the unit vectors
perpendicular to the periodic orbit at $\bm x^*(t)$ for $0\leq t < T$.
Note that these vectors are in the space spanned by the left
eigenvectors $\bm f_i(t)$ ($2\le i \le d$) of $\hat U_L(T+t,t)$,
whose eigenvalues are smaller than 1.
The variances in the normal space at $\bm x^*(t)$ for the steady
distribution are given by
\begin{eqnarray}
\lefteqn{
 \left<\Delta x_{\perp, i}(t)\Delta x_{\perp, j}(t)\right>_\infty
 \equiv 
\lim_{r\to\infty} {1\over\Omega}
 \left<\hat n_{\perp,i}(t)\right|\hat M(rT+t)\left|\hat n_{\perp,j}(t)\right>}
\nonumber\\
& = &
{1\over\Omega}\sum_{\ell, k=2}^d
    \left<\hat n_{\perp,i}(t) | e_\ell(t)\right>\left[
    {\lambda_\ell \lambda_k\over 1-\lambda_\ell \lambda_k}
\left<f_\ell\right|2\hat{\cal Q}_L(T)\left|f_k\right> 
+ \left<f_\ell\right|2\hat{\cal Q}_L(t)\left|f_k\right>
             \right]\left<e_k(t)|\hat n_{\perp,j}(t)\right>.
\label{Delta_x_perp}
\end{eqnarray}
In the case $|\lambda_2| \gg |\lambda_3|$, 
the terms with $\lambda_i$ for $i\ge 3$ can be neglected, thus
we obtain the approximate expression
\begin{align}
\lefteqn{
 \left<\Delta x_{\perp, i}(t)\Delta x_{\perp, j}(t)\right>_\infty
\approx }
\nonumber\\
& 
{1\over\Omega}
    \left<\hat n_{\perp,i}(t) | e_2(t)\right>\left[
    {\lambda_2^2 \over 1-\lambda_2^2}
\left<f_2\right|2\hat{\cal Q}_L(T)\left|f_2\right> 
+ \left<f_2\right|2\hat{\cal Q}_L(t)\left|f_2\right>
             \right]\left<e_2(t)|\hat n_{\perp,j}(t)\right>.
\end{align}
This means that the distribution extends in the direction of $\bm
e_2(t)$ projected on the normal space.  In the case of two-variable
system, i.e. $d=2$, Eq. (\ref{Delta_x_perp}) becomes simple because
there is only one normal vector $\hat{\bm n}(t)\parallel\bm f_2(t)$; the
variance perpendicular to the orbit is given by
\begin{equation}
 \left<\left(\Delta x_{\perp}^2(t)\right)\right>_\infty
=
{1\over\Omega}\,{1\over\left<f_2(t)|f_2(t)\right>}
\left[
    {\lambda_2^2 \over 1-\lambda_2^2}
    \left<f_2\right|2\hat{\cal Q}_L(T)\left|f_2\right> 
     + \left<f_2\right|2\hat{\cal Q}_L(t)\left|f_2\right>
\right].
\label{Delta_x^2_perp-2}
\end{equation}

\section{Numerical results for Brusselator}

\begin{table}
\begin{center}\begin{tabular}
{c@{\hspace{1ex}}|ccc| cc | cc}
\hline
$\rho$ & \multicolumn{3}{|c|}{Reactions}&
$\Delta X_\rho$ & $\Delta Y_\rho$ &transition rates $W_\rho$&$w_\rho$ 
\\\hline\hline
1 & & $\stackrel{k_1}{\longrightarrow}$ & $X$ & 1 & 0 & $k_1\Omega$ & $k_1$ 
\\
2 & $X$ & $\stackrel{k_2}{\longrightarrow}$ & $Y$ & $-1$ & 1 &$k_2 X$ & $k_2 x$ 
\\
3 & $2X+Y$ &$\stackrel{k_3}{\longrightarrow}$ & $3X$ & 1 & $-1$ & $\displaystyle k_3{X(X-1)Y\over \Omega^2}$ &
 $k_3 x^2 y$ 
\\
4 & $X$ & $\stackrel{k_4}{\longrightarrow}$ &  & $-1$ & 0 & $k_4X$ & $k_4 x$ 
\\
\hline
\end{tabular}\end{center}
\caption{Reaction table for Brusselator.}
\label{Brusselator}
\end{table} 

We numerically evaluate our formulas for
Brusselator\cite{Prigogine-1968}, i.e.  a simple model of the chemical
oscillation with two chemical species, $X$ and $Y$ (Table
\ref{Brusselator}).  From Eqs.(\ref{F_i}) and (\ref{Q_ij}), $F_i$ and
$Q_{ij}$ in the Hamiltonian are given by
\begin{align}
F_x & = k_1 - k_2 x + k_3 x^2 y - k_4 x ,
\\
F_y & = k_2 x - k_3 x^2 y,
\\
Q_{xx} & = {1\over 2}(k_1 + k_2 x + k_3 x^2 y + k_4 x)
\\
Q_{xy}=Q_{yx} & = -{1\over 2}(k_2 x + k_3 x^2 y)
\\
Q_{yy} & ={1\over 2}(k_2 x + k_3 x^2 y).
\end{align} 
The function $L_{ij}(t)$ defined by Eq.(\ref{L_ij}) are given by
\begin{align}
L_{xx}(t) & = -k_2 + 2 k_3\; x^*(t)\, y^*(t) - k_4 ,
\\
L_{xy}(t) & = k_3\; x^*(t)^2 ,
\\
L_{yx}(t) & = k_2 - 2 k_3\; x^*(t)\, y^*(t),
\\
L_{yy}(t) & = -k_3\; x^*(t)^2,
\end{align} 
where $(x^*(t),y^*(t))$ is a rate equation solution.
%
The covariance matrix $\hat M(t)$ of Eq.(\ref{hat_M}) is evaluated 
through Eq.(\ref{M_ij}) by
solving Eqs.(\ref{delta_x-eq}) and (\ref{delta_p-eq}) numerically with
proper initial conditions.
We also performed Monte Carlo simulations to simulate the original
master equation (\ref{master-eq}) using the event driven algorithm.

\begin{figure}
\includegraphics[width=8cm]{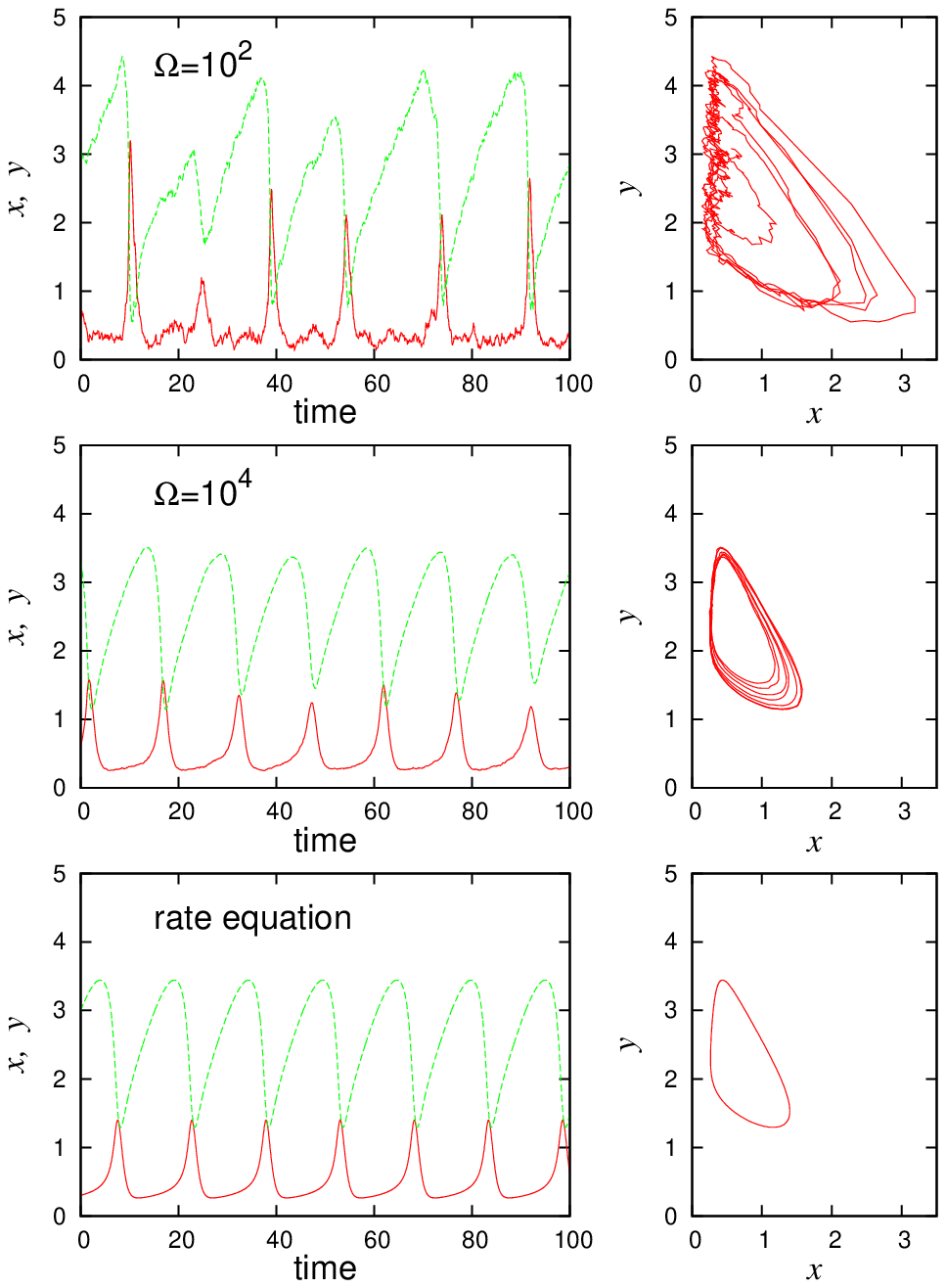}

\caption{Time sequences of Brusselator for $\Omega=100$ (top), 10,000
(middle), and the periodic rate equation solution (bottom).  In the left
panels, the concentrations $x$ and $y$ are plotted by the red and the
green lines respectively as a function of time.  In the right panels,
the trajectories are plotted in the $x-y$ plane.  The parameters for the
system are $k_1=0.5$, $k_2=1.5$, $k_3=1.0$, $k_4=1.0$, for which the
period of the rate equation solution is $T=15.1631$.}
\label{TimeSeq}
\end{figure}
\begin{figure}
\includegraphics[width=16cm]{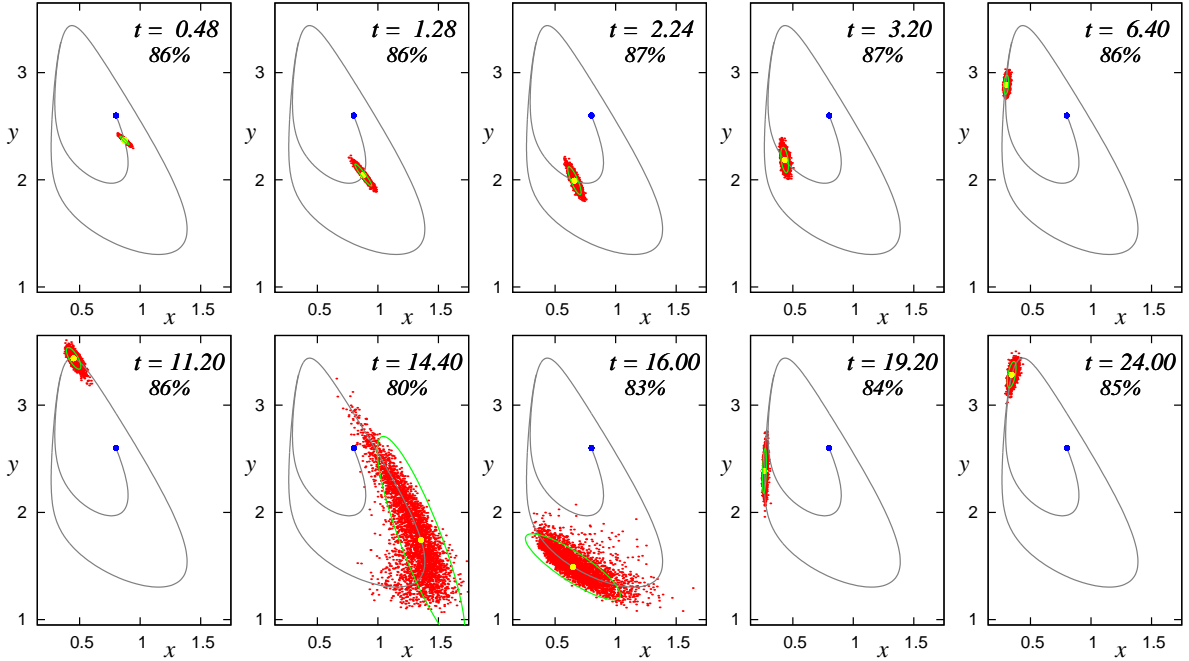}

\caption{Time development of distribution for Brusselator.  10,000
samples of Monte Carlo simulations are plotted by the red dots along with
the covariance matrix $\hat M$ 
estimated by
Eq.(\ref{M_ij}) ; $\hat M$'s are represented by the green ellipses given
by $\delta\bm x^T\hat M^{-1}\delta\bm x=4/\Omega$, where $\delta\bm
x^T\equiv (x-x^*(t),y-y^*(t))$.
The percentages of the samples that fall within the ellipses are shown
in each panel.
The gray curves represent the trajectory by the rate equation starting
from the initial point marked by the blue circles.  The system
parameters are $k_1=0.5$, $k_2=1.5$, $k_3=1.0$, $k_4=1.0$, and
$\Omega=10^4$.  The initial point is $(x_0^*, y_0^*)=(0.8,2.6)$.}
\label{trajectory-1}
\end{figure}
\begin{figure}
\includegraphics[width=8cm]{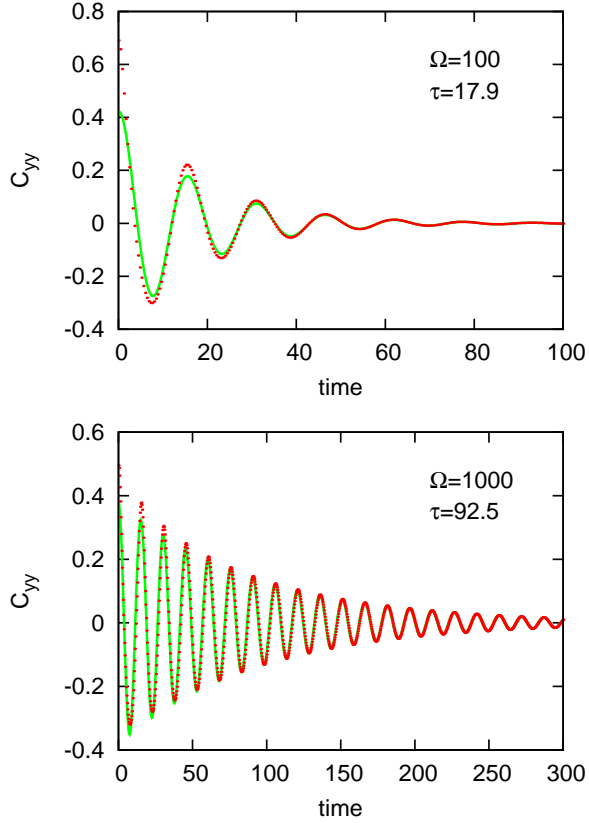}

\caption{Correlation functions $C_{yy}(t)$.  The dotted lines are the
results of Monte Carlo simulations and the green lines are fitting
functions by Eq.(\ref{fitting}) to estimate the relaxation time $\tau$.
$\Omega=$100 (the upper panel) and 1000 (the lower panel) and the rest
of the parameters are the same with those for Figs. \ref{TimeSeq} and
\ref{trajectory-1}.  } \label{correlation}
\end{figure}
\begin{figure}
{\includegraphics[width=8cm]{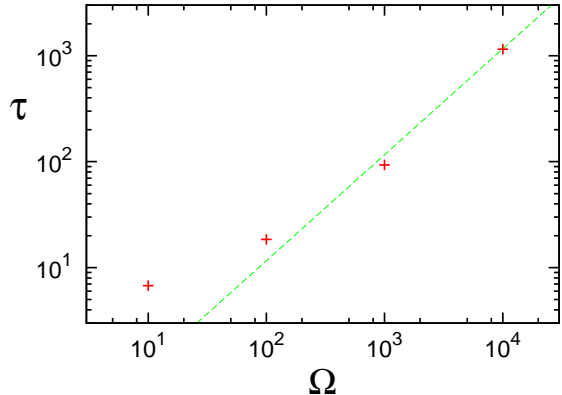}}

\caption{Correlation time vs. $\Omega$.  The red crosses shows the
relaxation times estimated by the Monte Carlo simulations, and the green
dashed line shows the line by Eq.(\ref{tau}).  The parameters are the
same with those for Figs. \ref{TimeSeq} and \ref{trajectory-1}, for
which $T=15.1631$ and 
$\left<f_1\right|2\hat{\cal Q}_L(T)\left|f_1\right>=1519.29$.
}
\label{corr-time}
\end{figure}
\begin{figure}
{\includegraphics[width=8cm]{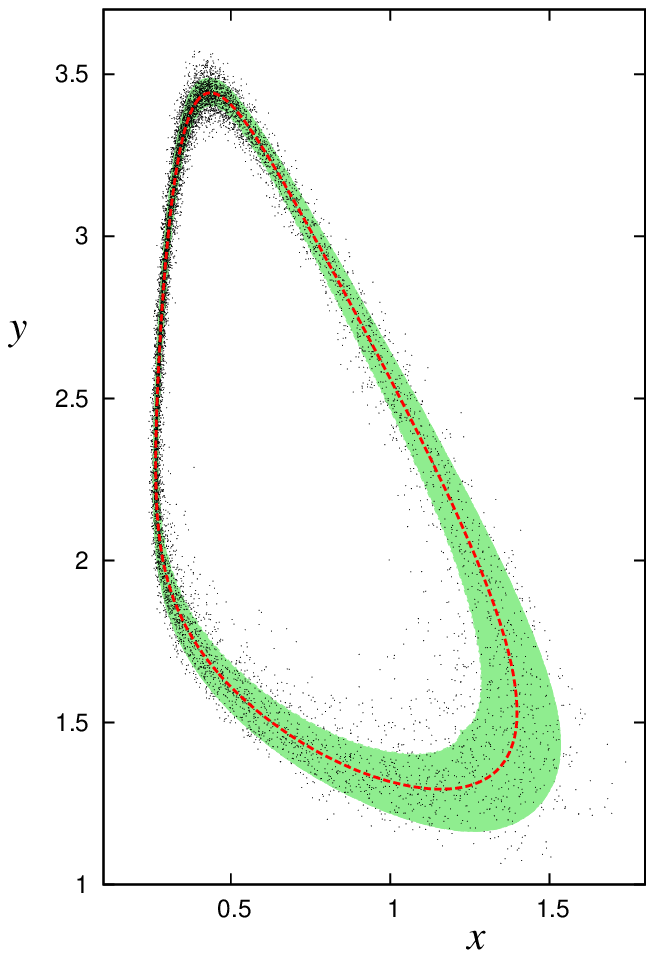}}

\caption{Steady state distribution.  The red dashed line represents the
periodic orbit by the rate equation, and the green area around it shows
the region within the distance of the square root of the local variance
Eq.(\ref{Delta_x^2_perp-2}) from the rate equation orbit. The dots show
10,000 samples by the Monte Carlo simulations.  $\Omega=10,000$ and
the rest of the parameters are the same with those for
Figs. \ref{TimeSeq} and \ref{trajectory-1}.  } \label{steady-dist}
\end{figure}

Figure \ref{TimeSeq} shows time sequences of the concentrations $x$ and
$y$ in sample trajectories for $\Omega = 100$ and 10,000 along with the
periodic rate equation solution.  The molecular fluctuation is large in
the smaller system.

Figure \ref{trajectory-1} shows the time development of the ensemble of
10,000 data points generated by Monte Carlo simulations along with the
ellipses given by the covariance matrix as
\begin{equation}
(\bm x-\bm x^*(t))^T\hat M(t)^{-1}(\bm x-\bm x^*(t))={4\over\Omega}
\end{equation}
with $\Omega=10^4$.  The initial point is marked by a blue circle and
the rate equation trajectory is shown by a gray curve in each plot.
One can see that the data point distributions obtained by Monte Carlo
simulations are fairly well represented by the ellipses.
The percentages of the samples that fall within the ellipse are shown in
each plots; Except for a few cases where the distribution is near the
curving parts of the trajectory, these percentages are close to 87\%,
i.e., the percentage for the two-dimensional Gaussian distribution.
The distribution extends initially transversely across the trajectory,
but eventually along the periodic orbit.  The distribution returns to
the Gaussian in the slow moving part of the trajectory even after it
deviates substantially from the Gaussian around the turning points in
the fast moving part.

The expression of period fluctuation (\ref{Delta_T^2}) is examined by
the relaxation time $\tau$ in the correlation function for the $y$
variable
\begin{equation}
 C_{yy}(t)\equiv \lim_{T_{\rm av}\to\infty}
 {1\over T_{\rm av}}\int_0^{T_{\rm av}}\Bigl[
  \left< y(t+t_0)y(t_0)\right>-\left<y(t+t_0)\right>\left<y(t_0)\right>
                                       \Bigr]dt_0 .
\label{C_yy}
\end{equation}
From simple calculation\cite{Gaspard-2002b}, the correlation function is
expected to decay as
\begin{equation}
   C_{yy}(t) \sim A e^{-t/\tau}\cos(\omega t) 
\label{fitting}
\end{equation}
with the relaxation time
\begin{equation}
 \tau = {T^3\over 2\pi^2 \left<\Delta T^2\right>}
=
 \Omega\,{T^3\over 
 2\pi^2 \left<f_1\right|2\hat{\cal Q}_L(T)\left|f_1\right>}.
\label{tau}
\end{equation}
Figure \ref{correlation} shows the correlation functions obtained by
Monte Carlo simulations (the red dotted lines) and the fitting functions
(\ref{fitting}) (the green lines) for $\Omega=100$ and 1,000.
The relaxation times obtained by the fitting are plotted in
Fig.\ref{corr-time} by the red crosses vs. $\Omega$ with the estimate by
Eq.(\ref{tau}) (the green dashed line).  They agree well for
$\Omega\gtrsim 100$.

The variance of the distribution width (\ref{Delta_x^2_perp-2})
perpendicular to the periodic orbit is shown by the green band in
Fig.\ref{steady-dist} with the dots that represents the ensemble of
10,000 samples by Monte Carlo simulations in the steady state.  
The width of the green band represents well the local width of the
distribution of the Monte Carlo simulations.

\section{Concluding remarks}

Using Hamilton-Jacobi method, we have derived a compact formula for the
conditional probability distribution by solving the chemical
Fokker-Planck equation within the Gaussian approximation.  The formula
is quite general and given for an arbitrary initial value of $\bm x_0$;
it can be applied not only to stationary states but also to oscillatory
states and transient processes.
In the case of stationary states, the formula reduces to the one
obtained by the linear noise
approximation\cite{vanKampen,Elf-2003,Scott-2007}.
In the case of oscillatory states, 
we have derived the
expressions for the phase diffusion along the stable periodic orbit and
for the steady distribution in the space perpendicular to the periodic
orbit.
We also have developed the method to evaluate these expressions
numerically.
Taking the Brusselator as an example, we have estimated the formulas
numerically and compared them with the results of the Monte Carlo
simulations; They agree quite well.

The present theory is based on the Hamilton-Jacobi formalism developed
by Gaspard\cite{Gaspard-2002b}, but the actual distribution functions
that we obtained are different; Gaspard expanded the distribution
function by the time variable, but we obtained the expression for the
distribution function by expanding by the $x$ variables
at a given time $t$.
Applying our formula for the periodic orbit, we obtain the expressions
for both phase diffusion along the orbit and the steady distribution in
the space perpendicular to the orbit.  The phase diffusion is directly
related to the temporal fluctuation, and the expression for the period
variance is derived.  Our expression for the period variance is
apparently different from the one by Gaspard\cite{Gaspard-2002b},
but they are shown to be equivalent.

Expressions equivalent to some of our results have been obtained using
slightly different method of
expansion\cite{Tomita-1974a,Vance-1996,Elf-2003}.  Although we kept only
the lowest order of $1/\Omega$ in the fluctuations with the Gaussian
approximation, our method is fairly straightforward and the final
expression is rather general and yet compact.  It also allows simple
numerical evaluation.


\appendix

\section{General Properties of Time evolution operator $\hat U_L(t,t_0)$}

In this appendix, various properties of the time evolution operator
$\hat U_L(t,t_0)$ are presented.
%
%
The time evolution operator (\ref{hat_U}) is defined as
\begin{align}
\hat U_L(t,t_0) & \equiv {\cal T}\exp\left[\int_{t_0}^t dt' \hat
L(t')\right] \nonumber \\ & \equiv 1 + \int_{t_0}^t dt_1 \hat L(t_1) +
\int_{t_0}^t dt_1 \int_{t_0}^{t_1} dt_2\, \hat L(t_1) \hat L(t_2) +
\cdots \nonumber \\ & = 1 + \int_{t_0}^t dt_1 \hat L(t_1) +
\int_{t_0}^t dt_1 \int_{t_1}^t dt_2\, \hat L(t_2) \hat L(t_1) + \cdots
,
\end{align} 
for the both cases of $t\ge t_0$ and $t\le t_0$.
The second equality gives the definition of the time ordering operator
${\cal T}$.  From this, it is easy to see that the operator satisfies
\begin{align}
{d\over dt}\hat U_L(t,t_0) & = \hat L(t)\hat U_L(t,t_0),
\\
{d\over dt_0}\hat U_L(t,t_0) & = -\hat U_L(t,t_0)\hat L(t_0),
\\
\hat U_L(t_0,t_0) & = 1,
\end{align}
and the following equalities hold:
\begin{align}
\hat U_L(t,t_0) & = \hat U_L(t,t_1)\hat U_L(t_1,t_0),
\\
\hat U_L^{-1}(t,t_0) & = \hat U_L(t_0,t),
\label{U-inv}
\\
\hat U_L^\dagger (t,t_0) & = \hat U_{-L^\dagger}(t_0,t),
\\
\left(\hat U_L^{-1}(t,t_0)\right)^\dagger 
& = \left(\hat U_L^\dagger(t,t_0)\right)^{-1} 
= \hat U_{-L^\dagger}(t,t_0).
\end{align}

\section{Right and Left Eigenvectors for $\hat U$}

For the time evolution operator $\hat U$ over the period $T$, the
expressions for the right and left eigenvectors, $\left|e_1\right>$ and
$\left<f_1\right|$, with the eigenvalue $\lambda_1=1$ can be obtained.


It is easy to see that the right eigenvector of $\lambda_1$ is given by
\begin{equation}
\left|e_1\right> = \left|F(\bm x_0^*)\right>
\label{e_1}
\end{equation} 
because from Eqs.(\ref{H-eq-x}) and (\ref{L_ij})
\begin{equation}
{d\over dt}F_i(\bm x^*(t)) = L_{ij}(t) F_j(\bm x^*(t)),
\end{equation} 
that gives
\begin{equation}
\left|F(\bm x_0^*)\right> = 
\left|F(\bm x^*(T))\right> = \hat U\left|F(\bm x_0^*)\right>.
\end{equation}


The corresponding left eigenvector $\left<f_1\right|$ can be also
obtained as follows.
Consider the periodic orbit $(\bm x^*(t;E),\bm p^*(t;E))$ outside the
$\bm p=0$ subspace, where the value of Hamiltonian $E$ is non-zero.
Suppose that it is an analytic function of $E$ with $(\bm x^*(t;0),\bm
p^*(t;0))=(\bm x^*(t),0)$, and  its period is given by $T(E)$ as a
function of $E$.

Let us define the deviations 
\begin{equation}
\delta\bm x^*(t) \equiv \bm x^*(t;\delta E)-\bm x^*(t), \quad
\delta\bm p^*(t) \equiv \bm p^*(t;\delta E)
\label{deviation}
\end{equation}
for small $\delta E$.  These satisfy Eqs.(\ref{delta_x-eq}) and
(\ref{delta_p-eq}), thus the formal solutions are given by
Eqs.(\ref{delta_x-sol0}) and (\ref{delta_p-sol0}).

Within the first order of $\delta E$, $\delta\bm p^*(0)$ is given by
\begin{align}
\delta\bm p^*(0) & = \bm p^*(T(\delta E);\delta E)
\approx \bm p^*(T(0);\delta E) = \delta\bm p^*(T),
\end{align}
where we have used $\bm p^*(t;0)=0$ and $T(0)=T$.
Using the formal solution (\ref{delta_p-sol0}),  the last expression is
written as
\begin{equation}
\delta\bm p^*(T) = \hat U_L^\dagger(0,T)\delta\bm p^*(0)
 = \left(\hat U^{-1}\right)^\dagger\delta\bm p^*(0),
\end{equation} 
which shows that $\left<\delta p^*(0)\right|$ is the left eigenvector
in the braket notation,
\begin{equation}
\left<\delta p^*(0)\right|\hat U = \left<\delta p^*(0)\right|.
\end{equation} 
From the normalization (\ref{norm}) with the right eigenvector
(\ref{e_1}), we have
\begin{equation}
\left< f_1\right| = {\left< \delta p^*(0)\right| \over\delta E}
=  {\partial \left<  p^*(0;E)\right| \over\partial E}\biggr|_{E=0}
\label{f_1}
\end{equation} 
because
\begin{equation}
\delta E = \left<\delta p^*(0)|F(\bm x_0^*)\right>
\end{equation} 
from Eq.(\ref{Hamiltonian}) within the lowest order of $\delta E$.

\section{Covariance matrix $\hat M(t+rT)$
after many periods of $T$}

We derive the expression for the covariance matrix $\hat M(rT+t)$ for a
large integer $r$ in the case of the periodic orbit $\bm x^*(t)$ with
the period $T$.  With the expression, we obtain Eq.(\ref{Delta_T^2}) for
the period fluctuation and Eq.(\ref{Delta_x_perp}) for the fluctuations
perpendicular to the orbit in the steady distribution.

For an integer $r$ and $0\leq t < T$, we can rewrite Eq.(\ref{hat_M}) as
\begin{align}
\lefteqn{
\hat M(rT+t) 
 = \hat U_L(rT+t,0)2\hat{\cal Q}_L(rT+t)\hat U_L^\dagger(rT+t,0)
}
\nonumber \\
& =  \hat U_L(rT+t,0)\left[
\int_0^{rT+t} dt' \hat U_L(0,t')2\hat Q(t')\hat U_L^\dagger(0,t')
\right]
\hat U_L^\dagger(rT+t,0)
\nonumber \\
& =
\hat U_L(rT+t,0)\Biggl[
\sum_{s=0}^{r-1}
\int_0^T dt' \hat U_L(0,sT+t')2\hat Q(t')\hat U_L^\dagger(0,sT+t')
\nonumber\\ & \qquad\qquad
+\int_0^t dt'\hat U_L(0,rT+t')2\hat Q(t')\hat U_L^\dagger(0,rT+t')
\Biggr]\hat U_L^\dagger(rT+t,0)
\nonumber \\
& =
\hat U_L(rT+t,0)\Biggl[
\sum_{s=0}^{r-1}
\hat U_L(0,sT)\left(
\int_0^T dt' \hat U_L(0,t')2\hat Q(t')\hat U_L^\dagger(0,t')\right)
\hat U_L^\dagger(0,sT)
\nonumber\\ & \qquad\qquad
+
\hat U_L(0,rT)\left(
\int_0^t dt'\hat U_L(0,t')2\hat Q(t')\hat U_L^\dagger(0,t')
\right)\hat U_L^\dagger(0,rT)
\Biggr]\hat U_L^\dagger(rT+t,0)
\nonumber \\
& =
\hat U_L(t,0)\Biggl[
 \sum_{s=0}^{r-1}
\hat U^{r-s}
\, 2\hat{\cal Q}_L(T)
\left(\hat U^{r-s}\right)^\dagger
+
2\hat{\cal Q}_L(t)
\Biggr]\hat U^\dagger_L(t,0)
\nonumber \\
& =
\hat U_L(t,0)\Biggl[ \sum_{s=1}^r 
   \hat U^s\, 2\hat{\cal Q}_L(T) \left(\hat U^s \right)^\dagger
+2\hat{\cal Q}_L(t)\Biggr]\hat U^\dagger_L(t,0),
\end{align} 
where we have used $\hat Q(t+T)=\hat Q(t)$.

For $t=0$, $\hat U_L(t,0)=1$ and $\hat{\cal Q}_L(t)=0$, thus
using the spectral representation
Eq.(\ref{U_spectral}) into this, we have
\begin{align}
{1\over r}\left<e_1\right| \hat M(rT)\left|e_1\right> 
& =
{1\over r}\sum_{s=1}^r 
\left< e_1\right|
  \left(\sum_i \left|e_i\right>\lambda_i^s\left<f_i\right|\right)
           2\hat{\cal Q}_L(T)
  \left(\sum_j \left|f_j\right>\lambda_j^s\left<e_j\right|\right)
\left|e_1\right>
\nonumber\\
& \rightarrow
\left<e_1 | e_1\right> 
  \left<f_1\right|2\hat{\cal Q}_L(T)\left|f_1\right>
     \left<e_1 | e_1\right>
\end{align} 
as $r\to\infty$ because $|\lambda_i|<1$ for $i\geq 2$.
From Eq.(\ref{Delta_T^2_0}) with Eq.(\ref{e_1}),
this gives 
\begin{equation}
\left<\Delta T^2\right> =
{1\over \Omega}\, \left<f_1\right| 2\hat{\cal Q}_L(T)\left|f_1\right>,
\label{Delta_T^2-app}
\end{equation} 
which is Eq.(\ref{Delta_T^2}).

The variances to the direction $\bm n_{\perp,i}(t)$ normal to the orbit
at $\bm x^*(t)$ can be calculated as
\begin{align}
\lefteqn{
\left<\Delta x_{\perp, i}(t)\Delta x_{\perp, j}(t)\right>
 =
{1\over \Omega}
\left<n_{\perp,i}(t)\right|\hat M(t+rT)\left|n_{\perp,j}(t)\right>
}\nonumber\\
\qquad & =
{1\over \Omega}\sum_{\ell,k= 2}^d\left<n_{\perp,i}(t)|e_\ell(t)\right>
\left[
\sum_{s=1}^r \lambda_\ell^s
     \left<f_\ell\right|2\hat{\cal Q}_L(T)\left|f_{k}\right>
             \lambda_{k}^s
+
   \left<f_\ell\right|2\hat{\cal Q}_L(t)\left|f_{k}\right>
\right]
\left<e_{k}(t)|n_{\perp,j}(t)\right>
\nonumber\\
& \stackrel{r\to\infty}{\longrightarrow}
{1\over \Omega}\sum_{\ell,k = 2}^d\left<n_{\perp,i}(t)|e_\ell(t)\right>
\left[
{ \lambda_\ell\lambda_{k}\over 1-\lambda_\ell\lambda_{k}}
     \left<f_\ell\right|2\hat{\cal Q}_L(T)\left|f_{k}\right>
+
   \left<f_\ell\right|2\hat{\cal Q}_L(t)\left|f_{k}\right>
\right]
\left<e_{k}(t)|n_{\perp,j}(t)\right>,
\label{Delta_x_perp-app}
\end{align} 
which is Eq.(\ref{Delta_x_perp}).
The matrix elements in
Eqs.(\ref{Delta_T^2-app}) and (\ref{Delta_x_perp-app}) can be
numerically estimated by Eq.(\ref{QL_ij}),
\begin{align}
\left<f_\ell\right|2\hat{\cal Q}_L(t)\left|f_k\right>
& = 
-\left<\delta p(t;f_\ell)|\delta_p x(t;f_k)\right>.
\end{align}

\section{Equivalence of Eq.(\ref{Delta_T^2}) 
with Gaspard's expression}

Gaspard\cite{Gaspard-2002a,Gaspard-2002b} has given the correlation
time in terms of the energy derivative of the period $T(E)$.  We show
his expression is equivalent with our expression
Eq.(\ref{Delta_T^2}).

As in Appendix B, the deviation of the periodic orbit $\delta\bm x^*(t)$
in Eq.(\ref{deviation}) at $t=0$ is expressed within the lowest order of
$\delta E$ as
\begin{align}
\delta\bm x^*(0) & =
\bm x^*(T(\delta E);\delta E) - \bm x^*(T(0);0)
\nonumber \\
& \approx \delta\bm x^*(T(0)) + \bm F(\bm x_0^*) T'(0) \delta E,
\label{C4}
\end{align}
where $T'(E)$ denotes the $E$ derivative of $T(E)$.
Again, using the formal solution (\ref{delta_x-sol0}) with the notation
(\ref{U}), we have
\begin{equation}
\delta\bm x^*(T) 
= \hat U\left(
     \delta\bm x^*(0)-2\hat{\cal Q}_L(T)\delta\bm p^*(0)\right)
\end{equation} 
By inserting this into Eq.(\ref{C4}) with Eqs.(\ref{e_1}) and (\ref{f_1}), we have
\begin{equation}
(1-\hat U)\left|\delta x^*(0)\right>
= \left(-\hat U 2\hat{\cal Q}_L(T)\left|f_1\right> + \left|e_1\right> T'(0)\right)\delta E
\end{equation}
in the braket notation.
Taking the inner product of this with $\left<f_1\right|$, 
i.e. the left eigenvector of $\hat U$ with the eigenvalue 1,
this gives
\begin{equation}
T'(0) = \left< f_1\right|2\hat{\cal Q}_L(T)\left|f_1\right>,
\end{equation} 
therefore, from Eq.(\ref{Delta_T^2}) we obtain
\begin{equation}
\left< \Delta T^2\right> 
    = {1\over\Omega}\,{\partial T\over\partial E}\Bigr|_{E=0}.
\end{equation}
With Eq.(\ref{tau}), this
 is equivalent to Gaspard's expression.


%

\end{document}